\def\BR{{\cal B}}

\def\totLumin{(7.56 \pm 0.15)} 
\def\totNtautau{(6.90 \pm 0.14) \times 10^6}

\def\threehpi0{\tau^- \to h^- h^+ h^- \pi^0 \nu_\tau}
\def\threepipi0{\tau^- \to \pi^- \pi^+ \pi^- \pi^0 \nu_\tau}
\def\op{\tau^- \to \pi^- \omega \nu_\tau}
\def\ok{\tau^- \to K^- \omega \nu_\tau}
\def\oh{\tau^- \to h^- \omega \nu_\tau}
\def\twopiKpi0{\tau^- \to K^- \pi^+ \pi^- \pi^0 \nu_\tau}
\def\twoKpipi0{\tau^- \to K^- K^+ \pi^- \pi^0 \nu_\tau}
\def\threeKpi0{\tau^- \to K^- K^+ K^- \pi^0 \nu_\tau}

\def\hpi0{\tau^- \to h^- \pi^0 \nu_\tau}

\def\wstwopiKpi0{\tau^- \to \pi^- K^+ \pi^- \pi^0 \nu_\tau}
\def\wstwoKpipi0{\tau^- \to K^- \pi^+ K^- \pi^0 \nu_\tau}

\def\fwstwopiKpi0{\pi^- K^+ \pi^- \pi^0}
\def\fwstwoKpipi0{K^- \pi^+ K^- \pi^0}

\def\fthreehpi0{h^- h^+ h^- \pi^0}
\def\fthreepipi0{\pi^- \pi^+ \pi^- \pi^0}
\def\fok{K^- \omega}
\def\ftwopiKpi0{K^- \pi^+ \pi^- \pi^0}
\def\ftwoKpipi0{K^- K^+ \pi^- \pi^0}
\def\fthreeKpi0{K^- K^+ K^- \pi^0}

\def\threepi{\pi^+ \pi^- \pi^0}

\newcommand{\prdd}[3]{Phys. Rev. D {\bf #1}, #3 (#2)}

\newcommand{\prll}[3]{Phys. Rev. Lett. {\bf #1}, #3 (#2)}
\newcommand{\plb}[3]{Phys. Lett. B {\bf #1}, #3 (#2)}
\newcommand{\epjc}[3]{Eur. Phys. J. C {\bf #1}, #3 (#2)}
\newcommand{\cpc}[3]{Comput. Phys. Commun. {\bf #1}, #3 (#2)}

\newcommand{\zphysc}[3]{Z. Phys. C {\bf #1}, #3 (#2)}

\def\etal{{\em et~al.}}
\def\ibid{{\em ibid.}}

\documentclass[aps,prd,preprint,superscriptaddress,tightenlines,nofootinbib,floatfix]{revtex4}

\usepackage{graphicx}
\usepackage{dcolumn}
\usepackage{bm}

\begin{document}

\preprint{CLNS 04/1899}       
\preprint{CLEO 04-17}         

\title{Study of $\tau$ Decays to Four-Hadron Final States with Kaons}

\author{K.~Arms}
\author{K.~K.~Gan}
\affiliation{Ohio State University, Columbus, Ohio 43210}
\author{H.~Severini}
\affiliation{University of Oklahoma, Norman, Oklahoma 73019}
\author{D.~M.~Asner}
\author{S.~A.~Dytman}
\author{W.~Love}
\author{S.~Mehrabyan}
\author{J.~A.~Mueller}
\author{V.~Savinov}
\affiliation{University of Pittsburgh, Pittsburgh, Pennsylvania 15260}
\author{Z.~Li}
\author{A.~Lopez}
\author{H.~Mendez}
\author{J.~Ramirez}
\affiliation{University of Puerto Rico, Mayaguez, Puerto Rico 00681}
% ...
\collaboration{CLEO Collaboration} 
\noaffiliation

\date{January 14, 2005}

\begin{abstract} 
The $\tau$ decays to four hadrons have been studied with the CLEO III detector at the Cornell Electron Storage Ring (CESR) using $\totLumin$ $fb^{-1}$
of data collected near the $\Upsilon(4S)$ resonance. We present the first statistically significant measurements of $\BR(\twopiKpi0, {\rm excluding}\ K^0) =
(7.4 \pm 0.8 \pm 1.1)\times 10^{-4}$ and $\BR(\twoKpipi0) = (5.5 \pm 1.4 \pm 1.2)\times10^{-5}$, including the first observation of
the decay $\ok$ with branching fraction, $(4.1 \pm 0.6 \pm 0.7)\times 10^{-4}$. We also publish the first upper limit for $\BR(\threeKpi0) <\
4.8\ (6.1) \times 10^{-6}$ at 90\% (95\%) confidence level (C.L.).
\end{abstract}

\maketitle

The decays of the $\tau$ lepton to final states with kaons provide a clean probe of the strange sector of the weak charged current. These high-mass
decays can also be used to improve constraints on the $\tau$ neutrino mass. Measurement of the branching fractions and spectral functions to strange
final states can be used to extract the strange quark mass and the Cabibbo-Kobayashi-Maskawa element $V_{us}$. Previous measurements of the decays
$\twopiKpi0$ and $\ftwoKpipi0$~\cite{conjugates,CLEO99,ALEPH98} had large statistical errors due to large $\tau$ migration backgrounds as a consequence of
limited particle identification capability; there are no measurements of the spectral functions. The decay $\twopiKpi0$ can proceed via the $\fok$
intermediate state. Using a vector meson dominance model as well as $SU(3)_f$ relations between $\rho$ and $\omega$ mesons, Li~\cite{li97}
predicts $\BR(\ok) = \BR(\tau^- \to K^- \rho^0 \nu_\tau)$. In this Letter, we present the first statistically significant measurements of the
branching fractions for the decays $\twopiKpi0 ({\rm ex.}\ K^0)$ and $\ftwoKpipi0$, as well as the first published upper limit for the decay $\threeKpi0$.
We also investigate the substructure of the decays $\twopiKpi0$ and $\ftwoKpipi0$, yielding the first observation and measurement of $\ok$.

The data used in this analysis were collected with the CLEO III detector~\cite{CLEO3det} at CESR near the center-of-mass
energy 10.58 GeV. The sample corresponds to an integrated luminosity of $\totLumin$ $fb^{-1}$ containing $\totNtautau$ $\tau$-pair events produced in
$e^+e^-$ collisions. The detector features a four-layer silicon strip vertex detector, a wire drift chamber, and a Ring Imaging Cherenkov (RICH) detector
that is critical for this analysis. The specific ionization loss ($dE/dx$) measured in the drift chamber is used to identify hadron species with a
resolution of about 6\%. The RICH detector~\cite{RICH,RICH2} surrounds the drift chamber and uses thin LiF radiators to generate Cherenkov photons from
the incident charged particles. Generated photons propagate through an expansion volume of gaseous nitrogen at atmospheric pressure, and are detected
by multiwire proportional chambers filled with a mixture of TEA and CH$_4$ gases. RICH particle identification (ID) is available within $|\cos\theta| < 0.83$,
where the polar angle, $\theta$, is with respect to the incident beam. The electromagnetic calorimeter surrounds the RICH detector and measures
the energy, position, and lateral shape of showers induced by charged and neutral particles. The calorimeter contains 7784 CsI(Tl) crystals arranged in a
barrel section ($|\cos\theta| < 0.83$) and two endcaps ($0.83 < |\cos\theta| < 0.95$). These components operate inside a 1.5 T solenoidal magnetic field.
A muon detection system surrounds the solenoid with iron absorber interspersed with wire chambers operated in proportional mode.

The $\tau^+ \tau^-$ candidate events must contain four well-reconstructed charged tracks with zero net charge. Each event is divided into two hemispheres
(tag and signal) using the plane perpendicular to the thrust axis~\cite{farhi77}. The thrust axis is determined using all charged tracks and photons. We select
events in a 1-vs-3 topology by requiring one and three tracks in the tag and signal hemispheres, respectively. 

We require the missing momentum of the event to be in the central region of the detector ($|\cos\theta_{miss}| < 0.85$) to suppress radiative Bhabha and
$\mu$-pair backgrounds. In order to diminish hadronic background while maintaining a high acceptance for $\tau$ decays, we require the tag hemisphere to
have invariant mass less than 1.2 GeV/$c^2$, and that of the signal hemisphere to be less than the $\tau$ lepton mass. Hemisphere masses are calculated
using all photons and charged tracks. Charged tracks in the signal hemisphere are assigned masses according to particle ID, while tracks in the tag
hemisphere are assigned the pion mass. In order to suppress two-photon backgrounds, we require that the total event visible energy be greater than 40\% of the
center-of-mass energy.

The momentum of tracks in the tag hemispehere must be greater than 100 MeV/$c$ and must point
into the central region of the detector, $|\cos\theta| < 0.90$. Charged pion and kaon track candidates in the signal hemisphere must have momentum greater
than 200 MeV/$c$ and satisfy  $|\cos\theta| < 0.8$ for improved particle ID performance. To reduce beam-gas and $\tau$-migration events with $K_S^0$, the
distance of closest approach of each track to the $e^+e^-$ interaction point (IP) must be within 5 mm transverse to the beam and 5 cm along the beam
direction. To further reduce $K_S^0$ background, we reject events containing a pair of tracks with detached vertex greater than 1 cm from the IP and
$\pi^-\pi^+$ mass consistent with the nominal $K_S^0$ mass. In order to reduce migration from events with photon conversion in the detector media as well
as Dalitz decays of $\pi^0$, we reject an event if any track in the signal hemisphere is identified as an electron (see below).

Photon candidates are defined as isolated energy clusters in the calorimeter with a photon-like lateral shower profile and energy deposition greater than
60 (100) MeV in the barrel (endcap) region of the detector. Candidate $\pi^0$'s are reconstructed from two-photon combinations using only photons in the
barrel section of the calorimeter. For the $\ok$ decay channel, a $\pi^0$ candidate must satisfy $-4 < S_{\gamma\gamma} < +3$, where $S_{\gamma\gamma} =
(M_{\gamma\gamma} - M_{\pi^0})/\sigma_{\gamma\gamma}$ ($\sigma_{\gamma\gamma}$ is the mass resolution calculated from the energy and angular resolution
of each photon).
\begin{figure}[b]
\includegraphics*[width=3.40in]{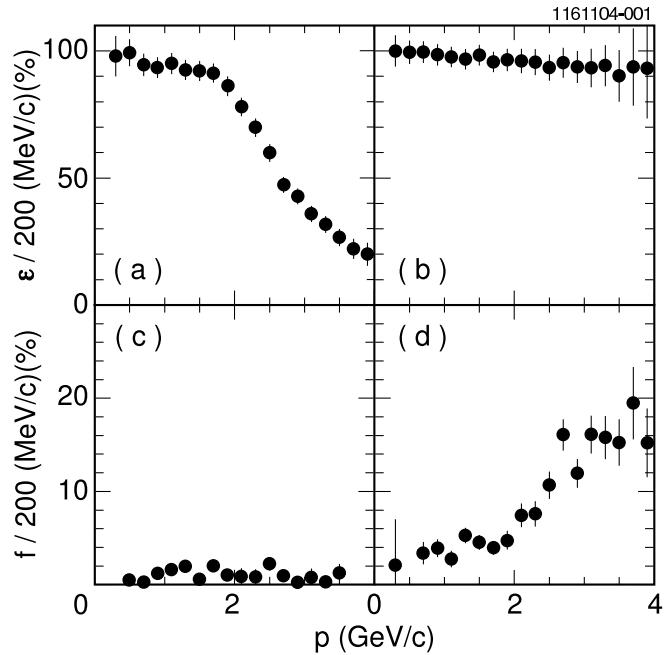}
\caption{Momentum dependent particle identification efficiency for kaons (a) and pions (b), and fake rates for $\pi$-as-$K$ (c) and  $K$-as-$\pi$ (d),
determined from a subsample of $D^{*+} \to D^0 \pi^+$ decays with $D^0 \to K^- \pi^+$ in the data. For the decay $\ok$, the  kaon detection efficiency
is $\sim$89\% and the pion fake rate from $\op$ is $\sim$1\%.}
\label{fig:class1}
\end{figure}

We do not explicitly veto events based on the photon multiplicity, thus allowing multiple $\pi^0$ candidates in an event in order to reduce
the dependence on Monte Carlo simulation of fake photons. However, we reject an event if any photon in the hemisphere not used in $\pi^0$ reconstruction has
energy in excess of 200 MeV, reducing feed-down from $\tau$ decays with more than one $\pi^0$ in the final state. To reduce combinatoric background, we also
require the angle of the $\pi^0$ candidate momentum in the $\tau$ center of mass frame relative to the $\tau$ momentum in the laboratory frame satisfy
$\cos\theta_{CM} > -0.75$. To boost a $\pi^0$ candidate to the $\tau$ rest frame, we approximate the $\tau$ energy as the beam energy, and the $\tau$
direction as along the visible momentum. Fake $\pi^0$'s show an enhancement at $\cos\theta_{CM} \simeq -1$, as opposed to a depletion for the signal.
In order to improve the $\omega$ mass resolution, the $\pi^0$ candidate is constrained to the nominal $\pi^0$ mass in calculating $M(\threepi)$.

We combine information from specific ionization loss ($dE/dx$) and from the RICH system to determine whether a track is identified as
a pion or kaon. The RICH detector response is distilled into a $\chi^2_i$ variable for each particle hypothesis
($i = \pi, K$). The value of $\chi^2_i$ is derived from the number of detected Cherenkov photons and their locations relative to the Cherenkov cone expected
for a particle with given momentum and mass. The $\chi^2_i$ from the RICH are combined with $\sigma_i$, the number of standard deviation of the measured
$dE/dx$ from expectation for the particle hypotheses, into the variable $\Delta\chi^2 = \chi^2_\pi - \chi^2_K + \sigma^2_\pi - \sigma^2_K$. A track is
identified as a pion (kaon) if it satisfies $\Delta\chi^2 < 0\ (> 10)$. The more stringent kaon requirement is necessary to reduce migration from the
dominant Cabibbo-favored $\tau$ decays with pions faking kaons. Efficiencies and fake rates are derived from kinematically selected $D^{*+} \to D^0 \pi^+$
decays with $D^0 \to K^- \pi^+$ in the data~\cite{wrongsign}. Figure~\ref{fig:class1} shows the efficiency and fake rate distributions for a subsample of the
data. The decreasing kaon identification efficiency for higher momenta is due to the stringent $\Delta\chi^2$ criteria, however few $\tau$ daughter tracks in
this sample have such high momentum. For candidate $\ok$ decays, we only identify the bachelor track and assume the tracks used to construct $\omega$
candidates are pions. For $\twopiKpi0\ ({\rm ex.}\ \omega)$ decays, we reject events with $M(\threepi)$ within 30 MeV/$c^2$ of the nominal $\omega$ mass.

The number of events in the decays $\twopiKpi0\ ({\rm ex.}\ \omega)$ and $\ftwoKpipi0$ are extracted by fitting the $S_{\gamma\gamma}$ distribution using
a Gaussian with a long low-mass tail over a polynomial background, in which the shape of the $\pi^0$ signal is constrained to the signal Monte Carlo
expectation. For the decay $\threeKpi0$, we simply require $-4 < S_{\gamma\gamma} < +3$; no events are observed in this window. The yield for $\ok$
is extracted by fitting the $M(\threepi)$ distribution using a Breit-Wigner lineshape~\cite{BW-shape} convoluted with a Gaussian resolution function over
a polynomial background.
\begin{table*}[b]
\begin{centering}
\vspace{0.1in}
\begin{tabular}{lccccc}
\hline
\hline
& \vspace{-0.10in} &&&\\
Channel                               \vspace{0.00in} \makebox[0.1in] & yield        \makebox[0.1in] & $\tau$ migration \makebox[0.1in] & $q\bar{q}$ bg \makebox[0.1in] & $\epsilon$ (\%) & $\BR$ ($\times 10^{-4}$) \\
\hline
& \vspace{-0.10in} &&&\\
$\ftwopiKpi0$ $({\rm ex.}\ K^0,\omega$) \vspace{0.05in} \makebox[0.1in] & $833 \pm 36$ \makebox[0.1in] & $434 \pm 14$     \makebox[0.1in] & $153 \pm 25$  \makebox[0.1in] & $5.68 \pm 0.17$ \makebox[0.1in] & $3.7 \pm 0.5 \pm 0.8$ \\
$\fok$                                \vspace{0.05in} \makebox[0.1in] & $500 \pm 35$ \makebox[0.1in] & $194 \pm 12$     \makebox[0.1in] & ~$64 \pm 20$  \makebox[0.1in] & $5.61 \pm 0.09$ \makebox[0.1in] & ~$4.1 \pm 0.6 \pm 0.7$ \\
$\ftwoKpipi0$                         \vspace{0.05in} \makebox[0.1in] & ~$48 \pm 9$  \makebox[0.1in] & ~~$1 \pm 1$      \makebox[0.1in] & ~~$9 \pm 7$   \makebox[0.1in] & $5.89 \pm 0.12$ \makebox[0.1in] & ~$0.55 \pm 0.14 \pm 0.12$ \\
$\fthreeKpi0$                         \vspace{0.00in} \makebox[0.1in] & $0$          \makebox[0.1in] & $0$              \makebox[0.1in] & $0$           \makebox[0.1in] & $4.36 \pm 0.10$ \makebox[0.1in] & $<\ 0.048\ (0.061)$ \\
\hline
\hline
\end{tabular}
\caption[]{Yields, backgrounds, efficiencies, and branching fraction measurements. The branching fraction for $\threeKpi0$ corresponds to the 90\%
(95\%) C.L. upper limit~\cite{Feldman98}. The second error is systematic.}
\label{tab:totresults}
\end{centering}
\end{table*}

Hadronic backgrounds are calculated empirically using a sample of high-mass tagged events assuming the two jets fragment independently. The hadronic background
calculation has been verified by finding consistent branching fraction measurements using a lepton ($e$ or $\mu$) tagged sample in which hadronic backgrounds are
greatly suppressed. An electron candidate must have specific ionization loss consistent with that expected for an electron and the ratio of shower energy to
momentum, $0.85 < E_{sh}/p < 1.1$. A muon candidate must penetrate at least three (five) absorption lengths of iron for track momentum less (greater) than 2.0 GeV/$c$.
\begin{figure}[b]
\includegraphics*[width=3.20in]{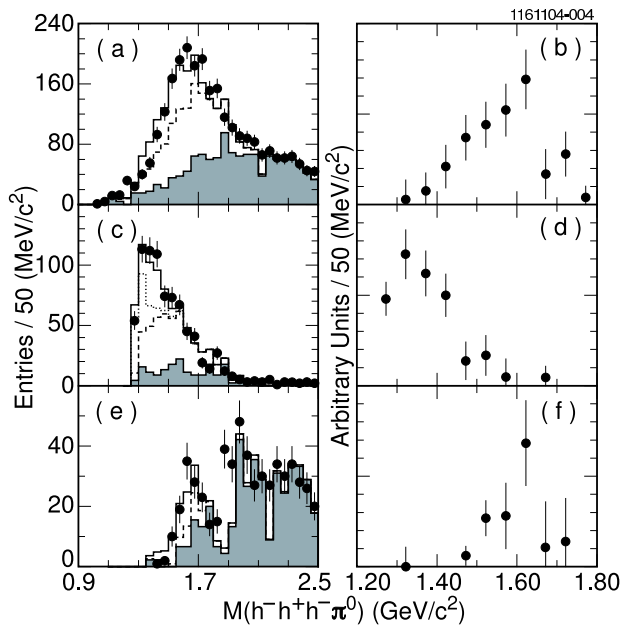}
\caption{Four-hadron invariant mass spectra for $\twopiKpi0\ ({\rm ex.}\ \omega)$ (a), $\fok$ (c), and $\ftwoKpipi0$ (e) events. The histograms
show the expectations, including $\tau$-migration Monte Carlo (dashed) and hadronic (shaded) background events. The dotted histogram in (c) shows
the contribution of $K_1(1270)$. The reproduction of the spectra above the $\tau$ mass by the high-mass tagged sample indicates the validity of the
hadronic background estimate. The distributions in (b,d,f) show the corresponding background- and efficiency-corrected spectra. Data points in (a)
below 1.3 GeV/$c^2$ are consistent with the background expectation and also below the production threshold in the $K^-a_1^0$ model, hence are not
displayed in (b).}
\label{fig:m4h-all}
\end{figure}
\begin{figure}[p]
\includegraphics*[width=3.20in]{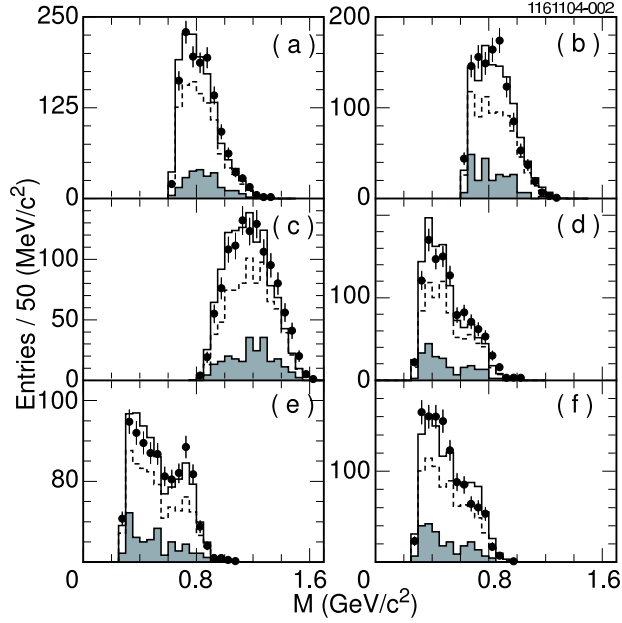}
\vspace{-0.15in}
\caption{Two- and three-hadron invariant mass spectra (a) $K^-\pi^+$, (b)$K^-\pi^0$, (c) $K^-\pi^+\pi^-$, (d) $\pi^-\pi^+$, (e) $\pi^+\pi^0$, and (f)
$\pi^-\pi^0$, in non-$\omega$ resonant $\twopiKpi0$ events. The histograms show the expectation including $\tau$-migration Monte Carlo events (dashed)
and hadronic background (shaded).} 
\label{fig:Kpipipi0-plots}
\end{figure}
\begin{figure}[p]
\includegraphics*[width=3.20in]{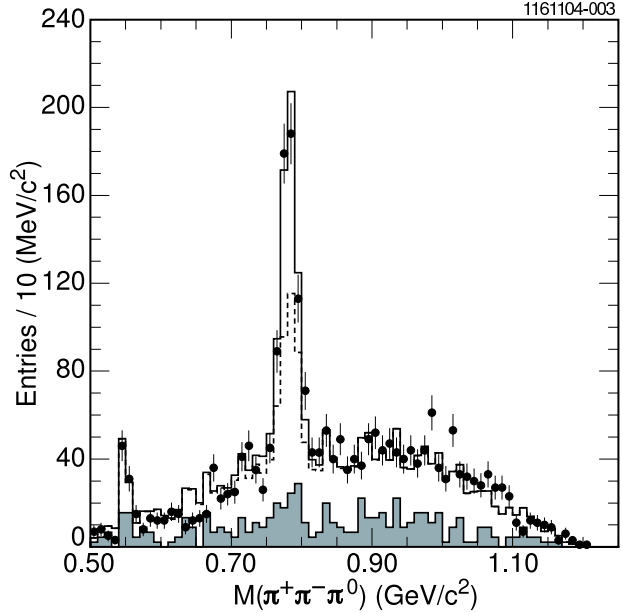}
\vspace{-0.15in}
\caption{$\threepi$ invariant mass spectrum in $\twopiKpi0$ events. The histogram shows the expectation using an equal mixture of $K_1(1400)$
and $K_1(1270)$, including $\tau$-migration Monte Carlo events (dashed) and hadronic background (shaded).} 
\label{fig:momega-ok}
\end{figure}

Efficiencies and $\tau$-migration backgrounds are estimated using Monte Carlo events generated from the {\verb$KORALB-TAUOLA$}
program~\cite{jadach85} with the detector response simulated by the {\verb$GEANT$} program~\cite{brun87}. We model the decay $\twopiKpi0\ ({\rm ex.}\ \omega) $ with the 
$K^-a_1^0$ intermediate state and the decays $\twoKpipi0$ and $\fthreeKpi0$ via phase space. The decay $\ok$ is modelled with an equal mixture
of $K_1(1270)$ and $K_1(1400)$ resonances. The signal, background, and detection efficiencies are summarized in Table~\ref{tab:totresults}. The
largest sources of $\tau$-migration backgrounds are from channels with a charged pion misidentified as a kaon.

We have investigated the hadronic mass spectra to search for substructure. Figure~\ref{fig:m4h-all}(a, c, e) shows the $h^-h^+h^-\pi^0$
invariant mass spectra for $\twopiKpi0\ ({\rm ex.}\ \omega)$, $\fok$, and $\ftwoKpipi0$ events. The background- and efficiency-corrected spectra
shown in Fig.~\ref{fig:m4h-all}(b, d, f) may be used to extract spectral functions for these final states. The two- and three-hadron substructure for
$\twopiKpi0\ ({\rm ex.}\ \omega)$ decays are shown in Fig.~\ref{fig:Kpipipi0-plots}.  There is no evidence for $K_1$, $\rho^0$, or $K^*$.
However, the $M(\pi^+\pi^0)$ spectrum (Fig.~\ref{fig:Kpipipi0-plots}(e)) indicates the presence of $\rho^+$ consistent with the prediction of the
$K^-a_1^0$ model; the observed $M(\pi^-\pi^0)$ spectrum (Fig.~\ref{fig:Kpipipi0-plots}(f)) is not inconsistent with the $K^-a_1^0$ model,
$\chi^2/d.o.f. = 24/(13-1)$ (C.L. = 2.3\%), though there is no clear indication of a $\rho^-$. The distribution of $M(\threepi)$ in
Fig.~\ref{fig:momega-ok} clearly shows an $\omega$ signal, as well as a small $\eta$ meson peak consistent with expectations for the decay
$\tau^- \to K^- \eta \nu_\tau$~\cite{CLEO96,ALEPH97}. This is the first observation of the decay $\ok$. The Monte Carlo model with an equal
mixture of $K_1(1270)$ and $K_1(1400)$ resonances describe the $K^-\omega$ mass distribution well, though we do not attempt to extract the relative
fraction of the two $K_1$ intermediate states. There is no indication of $K^*$ or $\rho$ resonances in the decay $\twoKpipi0$.

Most sources of systematic uncertainty affect these channels similarly. These include the uncertainty in integrated luminosity (2\%), $\tau$-pair cross-section
(1\%), charged track reconstruction (0.5\% per track), 1-prong tag branching fraction (0.1\%), electron veto (1.5\%), photon veto (3\%), $\pi^0$ detection
(8.8\%), particle ID efficiency (1.1--3.0\%), $\tau$-migration background due to particle misidentification (8.1--15.6\%) and limited
Monte Carlo statistics (2.6--5.7\%), hadronic background estimate (8.3--18.4\%), and detection efficiency (1.7--3.0\%) due to limited Monte Carlo
statistics. The $\pi^0$ detection efficiency has been calibrated using the high statistics decay $\hpi0$. The $\ok$ channel has additional uncertainty due to the branching
fraction of $\omega \to \threepi$ (0.8\%), model dependence from use of either $K_1$ model alone (2.6\%), and from variation of the polynomial background
parameterization in fitting the $\omega$ signal (2.5\%).

The branching fractions with systematic errors are summarized in Table~\ref{tab:totresults}. The branching fractions for $\BR(\twopiKpi0,\ {\rm ex.}\ K^0,\
\omega)$ and $\BR(\ok)\times\BR(\omega \to \threepi)$ can be combined to yield $\BR(\twopiKpi0,\ {\rm ex.}\ K^0) = (7.4 \pm 0.8 \pm 1.1) \times
10^{-4}$~\cite{K3piRatio}, which is significantly more precise than previous measurements~\cite{CLEO99,ALEPH98}. Comparing the measurement of $\ok$ with
the previous measurement of $\tau^- \to K^- \rho^0 \nu_\tau$~\cite{PDG04}, the ratio of branching fractions is significantly below Li's
prediction~\cite{li97,KRhoPred}. The measurement of $\BR(\twoKpipi0)$ is almost an order of magnitude smaller than previous results~\cite{CLEO99,ALEPH98} which were
based on samples with significantly larger $\tau$ migration backgrounds due to limited particle identification capability. The result on $\threeKpi0$
corresponds to the first published upper limit on this decay.

We gratefully acknowledge the effort of the CESR staff in providing us with excellent luminosity and running conditions.
This work was supported by the National Science Foundation and the U.S. Department of Energy.


\begin{thebibliography}{99}

\bibitem{conjugates}
Charge conjugate states are implied throughout this Letter.

\bibitem{CLEO99}
CLEO Collaboration, S.J.~Richichi~\etal, \prdd{60}{1999}{112002}.

\bibitem{ALEPH98}
ALEPH Collaboration, R.~Barate~\etal, \epjc{1}{1998}{65}.

\bibitem{li97}
Bing~An~Li, \prdd{55}{1997}{1436}; The ratio $\frac{\BR(\ok)}{\BR(\tau^- \to K^- \rho^0 \nu_\tau)}$ should be $1$ instead of $\frac{1}{3}$ in the paper
according to private communication with the author.

\bibitem{CLEO3det} 
CLEO Collaboration, Y.~Kubota {\it et~al.}, {Nucl. Instrum. Meth. A} \textbf{320}, {66} ({1992}); D.~Peterson \etal,
{Nucl. Instrum. Meth. A} \textbf{478}, {142} ({2002}).

\bibitem{RICH}
T.~Coan, {Nucl. Instrum. Meth. A} \textbf{379}, {448} ({1996}).

\bibitem{RICH2}
CLEO Collaboration, M.~Artuso \etal, {Nucl. Instrum. Meth. A} \textbf{502}, {91} ({2003}).

\bibitem{farhi77}
E.~Farhi, \prll{39}{1977}{1587}.

\bibitem{wrongsign}
We have measured the ``wrong-sign'' decays, $\wstwopiKpi0$ and $\fwstwoKpipi0$, observing no excess of events,
indicating a reliable estimate of the fake rate.

\bibitem{BW-shape}
The spin-dependent Breit-Wigner shape used is of the form $f(s) \propto \frac{\sqrt{s}m\Gamma}{(s-m^2)^2+m^2\Gamma^2}$.

\bibitem{jadach85}
S.~Jadach and Z.~Was, \cpc{36}{1985}{191}; {\bf 64}, 267 (1991); S.~Jadach, 
J.~H.~Kuhn, and Z.~Was, \ibid~{\bf 64}, 275 (1991).

\bibitem{brun87}
R.~Brun \etal, CERN Report No. CERN-DD/EE/84-1, 1987 (unpublished).

\bibitem{Feldman98}
G.J.~Feldman and R.D.~Cousins, \prdd{57}{1998}{3873}.

\bibitem{CLEO96}
CLEO Collaboration, J.E.~Bartelt~\etal, \prll{76}{1996}{4119}.

\bibitem{ALEPH97}
ALEPH Collaboration, D.~Buskulic~\etal, \zphysc{74}{1997}{263}.

\bibitem{K3piRatio}
The ratio $\BR(\ok)/\BR(\twopiKpi0,\ {\rm ex.}\ K^0) = (0.554 \pm 0.056 \pm 0.063)$ is a bit higher than but consistent
with the ratio  $\BR(\oh)/\BR(\threehpi0,\ {\rm ex.}\ K^0) = (0.446 \pm 0.015)$~\cite{PDG04}, suggesting a similar
hadronization pattern in $\twopiKpi0$ and $\threepipi0$ decays.

\bibitem{PDG04}
Particle Data Group, S.~Eidelman~\etal, \plb{592}{2004}{1}.

\bibitem{KRhoPred}
The absolute prediction for $\BR(\tau^- \to K^- \rho^0 \nu_\tau)$ is also significantly lower
than the experimental result.

\end{thebibliography}
\end{document}